\documentclass[10pt,a4paper]{article}
\usepackage{mypackages}
\usepackage{helvet}

\makeatletter
\def\input@path{
	{FigTex}
}
\makeatother
\graphicspath{{FigTex}}

\usepackage[sorting=none]{biblatex}
\bibliography{References.bib}

\begin{document}

% ======= Titre =======
\papertitle{Impact of Torque Ripple Induced Angular Oscillations on Maxwell Pressure Harmonics in Permanent-Magnet Synchronous Machines}

% ======= Auteurs =======
\paperauthors{Thomas Poupon\textsuperscript{1,2}, Arthur Givois\textsuperscript{1}, Vincent Lanfranchi\textsuperscript{1}, Jean-Daniel Chazot\textsuperscript{1}, Tahar Hamiti\textsuperscript{2}}

% ======= Affiliations =======
\paperaffiliations{
\textsuperscript{1}Roberval, Université de Technologie de Compiègne (UTC), Centre de recherche Royallieu, CS 60319 - 60203, Compiègne, France\\
\textsuperscript{2}Nidec-PSA-Emotors, 212, Bd Pelletier, 78 955, Carrières-Sous-Poissy, France
}

% ======= Résumé =======

\vspace{0.5cm}
\begin{multicols}{2}
\textbf{Abstract:}
Torque ripple in electrical machines is a well-known source of noise and vibration.
While numerous studies focus on control strategies to mitigate this ripple,
the feedback mechanism by which the resulting mechanical oscillations influence the electromagnetic forces remains less explored.
This paper presents a detailed investigation into the impact of torque ripple on the rotor angular dynamics
and the subsequent generation of new harmonics in the Maxwell pressure distribution. We first develop an analytical model for a rigid rotor subjected to a single torque harmonic,
revealing the inherently non-linear nature of the angular response.
This analysis demonstrates the creation of new harmonics in the rotor motion and a slight reduction in the mean rotational speed.
The study is then extended to a flexible rotor model incorporating torsional modes, solved using the Finite Element Method.
Finally, we introduce a strongly coupled magneto-mechanical model, based on the subdomain method,
to quantify how these rotor oscillations modulate the air-gap magnetic field.
The results show a new modulated spatio-temporal harmonic distribution in the Maxwell pressure spectrum, which is not predicted by conventional uncoupled models.
These findings are crucial for high-fidelity Noise, Vibration, and Harshness (NVH) analysis, as these additional pressure waves can excite stator structural modes.
\noindent\textbf{Keywords:} PMSM, Torque Ripple, Maxwell Pressure, Magneto-Mechanical Coupling, NVH

% ======= Corps du document =======
\vspace{-0.1cm}
\section*{Introduction}

In the design and operation of permanent-magnet synchronous machines (PMSMs),
torque ripple is a primary concern, directly contributing to undesirable noise and vibrations \cite{Timar_1989}.
The structural dynamics of the stator are particularly critical, as it is the main body responsible for radiating acoustic noise,
making its modeling and design key to vibration mitigation \cite{MILLITHALER2015185, Chauvicourt_2019}.
A significant body of literature is therefore dedicated to developing control strategies aimed at minimizing this ripple,
focusing on aspects such as optimal winding design, rotor shaping, or advanced current injection techniques \cite{Wu_2021, Thapa_2022, benharref2024full}.
These studies typically treat the electromagnetic torque as the final output of the system to be controlled, without deeply considering the mechanical system reaction to it.

However, the electromagnetic torque acts as an excitation on the rotor, inducing torsional oscillations and fluctuations in its rotational speed.
This dynamic response means the rotor angular position is not a simple linear function of time. This modulation of the angular position, in turn,
alters the relative position between the rotor magnetic field and the stator magnetomotive force, thereby modifying the air-gap flux density distribution.
Consequently, the very Maxwell pressures that generate the torque are themselves affected by the mechanical response,
creating a fully coupled electro-magneto-mechanical phenomenon.
As demonstrated in an asynchronous machine \cite{Li_2021}, modeling this interaction by considering the instantaneous angular speed and the effect of rotor displacement
is crucial for accurately analyzing the machine dynamic behavior.

This paper addresses a notable gap in the existing literature: the analysis of the closed-loop impact of torque ripple,
from the electromagnetic source to the mechanical response and back to the electromagnetic forces.
We aim to demonstrate and quantify how angular oscillations induced by torque ripple create new, and potentially significant,
spatio-temporal harmonics in the Maxwell pressure exerted on the stator. This investigation is structured as follows:
Section 1 details the non-linear angular response of the rotor, first with a rigid model to gain analytical insight,
and then with a flexible model. Section 2 presents the electromagnetic model based on the subdomain method,
formulated to account for these dynamic angular variations. Finally, Section 3 presents the results from a strongly coupled magneto-mechanical simulation,
highlighting the effects on both the rotor dynamics and the stator pressure distribution.

\section{Rotor Dynamic Response to Angular Excitation}

To understand the core phenomena, we first analyze the rotor dynamic response to torque ripple, isolating it from the full electromagnetic coupling.

\subsection{Rigid Rotor Model: The Non-Linear Response}

Let us consider a rigid shaft with an inertia $J$ rotating at a nominal angular velocity $\Omega_0$. The equation of motion is given by:
\begin{equation}
  J \ddot{\alpha} = C_{\text{mag}} - C_{\text{load}}
\end{equation}
where $\alpha$ is the angular position, $C_{\text{mag}}$ is the electromagnetic torque, and $C_{\text{load}}$ is the resistive load torque. The electromagnetic torque can be expressed as a Fourier series:
\begin{equation}
  C_{\text{mag}} = C_{0} + \sum_p^{\infty}(C_c \cos(p\alpha) + C_s \sin(p\alpha))
\end{equation}

To isolate the influence of a single torque ripple component, we consider an excitation of the form $C \cos(p \alpha)$, where $p$ is the harmonic order. We further assume that the mean electromagnetic torque $C_0$ perfectly balances the load torque $C_{\text{load}}$. The equation of motion simplifies to the following non-linear equation:
\begin{equation}
  J \ddot{\alpha} = C \cos(p \alpha)
\end{equation}
Multiplying each side by $\dot{\alpha}$ and integrating with respect to time yields the conservation of energy:
\begin{equation}
  \frac{1}{2} J \dot{\alpha}^2 = \frac{C}{p} \sin(p \alpha) + E_0
\end{equation}
where $E_0 = \frac{1}{2} J \Omega_0^2$ is the initial kinetic energy, with $\Omega_0$ being the initial angular velocity.

Isolating $\dot{\alpha}$ gives the first-order differential equation:
\begin{equation}
  \dot{\alpha} = \frac{d \alpha}{d t} = \sqrt{\frac{2 C}{J p} \sin(p \alpha) + \Omega_0^2}
\end{equation}
Separating variables allows us to express the time $t$ as a function of the angle $\alpha$ in an integral form:
\begin{equation}
  t(\alpha) = \int_{\alpha_0}^\alpha \frac{d\alpha'}{\sqrt{\frac{2 C}{J p} \sin(p \alpha') + \Omega_0^2}}
\end{equation}
This integral is a form of the incomplete elliptic integral of the first kind, $F(\phi,m)$. Inverting this relationship to find $\alpha(t)$ involves the Jacobi elliptic functions.
The exact solution for $\alpha(t)$ is given by:
\begin{equation}
  \alpha(t) = \frac{2}{p} \left[ \frac{\pi}{4} - \operatorname{am}\bigl(- E \cdot (t - t_0), m\bigr) \right]
\label{eq:alpha_t_simplified}
\end{equation}
where $\operatorname{am}(u,m)$ is the Jacobi amplitude function, the characteristic frequency $E$, the time shift $t_0$, and the elliptic modulus $m$ are defined as:
\begin{gather}
  E = \frac{p}{2} \sqrt{\frac{J p \Omega_0^2 + 2 C}{J p}} \label{eq:omega_c} \\
  t_0 = \frac{\pi}{2 p \Omega_0} \label{eq:t0} \\
  m = \frac{4 C}{J p \Omega_0^2 + 2 C} \label{eq:elliptic_m}
\end{gather}

While exact, this form is not intuitive. Since the torque ripple amplitude $C$ is typically small compared to the rotational kinetic energy term $J \Omega_0^2$, a Taylor series expansion in powers of $C$ provides a more practical and insightful expression for the angular position:
\begin{small}
\begin{align}
&\alpha(t) \approx{} \Omega_1 t %\notag \\
                   - \frac{C}{J p^2 \Omega_0^2} \cos(p \Omega_1 t)% \notag \\
                   - \frac{C^2}{8 J^2 \Omega_0^4 p^3} \bigl(\sin(2 p \Omega_1 t) - 3 \pi \bigr) \notag \\
                  & - \frac{C^3}{24 J^3 \Omega_0^6 p^4} \Bigl( 30 \cos(p \Omega_1 t)  - 2 \cos^3(p \Omega_1 t) - 9 \pi \sin(p \Omega_1 t) \Bigr)% \notag \\  \qquad
                   \notag \\
                  &+ \mathcal{O}(C^4)
\label{eq:alpha_taylor_split}
\end{align}
\end{small}

where the corrected mean angular velocity $\Omega_1$ is itself a function of $C$:
\begin{equation}
\Omega_1 = \Omega_0 
- \frac{3 C^2}{4 J^2 \Omega_0^3 p^2}
- \frac{69 C^4}{64 J^4 \Omega_0^7 p^4}
- \frac{315 C^6}{128 J^6 \Omega_0^{11} p^6} + \mathcal{O}(C^8).
\end{equation}

This analytical development leads to three key conclusions:\\
\textbf{Mean Speed Reduction:} The torque ripple induces a second-order quadratic decrease in the mean rotational speed, proportional to $C^2$. This implies that torque ripple leads to a slight decrease in the machine efficiency.\\% and can cause an order shift in FFT analysis.\\
\textbf{Generation of harmonics:} The non-linear nature of the response generates motion components at multiples of the initial excitation order ($2p$, $3p$, etc.), even though the input torque contained only order $p$. This is a fundamental departure from a linear, frequency-domain analysis.\\
\textbf{Low-Frequency Dominance:} The amplitudes of the oscillation terms are proportional to $1/p^2$ and $1/\Omega_0^2$. This indicates that the effect of torque ripple on the angular dynamics is most pronounced for low-order harmonics and at lower rotational speeds.

\subsection{Flexible Rotor Model with Torsional Modes}

The rigid body assumption is an idealization. Real-world rotors exhibit torsional flexibility, which can lead to resonant amplification of torque ripple at specific frequencies. To capture these effects, the rotor is modeled as a one-dimensional beam undergoing torsional deformation. The governing partial differential equation is solved using the Finite Element Method (FEM), resulting in a set of torsional modes and natural frequencies. The parameters used for this model are detailed in Table \ref{tab:rotor_params}.

\begin{table}[H]
\centering
\caption{Dimensions and mechanical properties of the rotor.}
\label{tab:rotor_params}
\begin{tabular}{l r}%[table-format=3.1]
\hline
\textbf{Parameter} & \textbf{Value} \\
\hline
Length $L$ [\si{\mm}]                     & 350    \\
Shaft diameter $D$ [\si{\mm}]             & 40.3   \\
Mass density $\rho_{\text{rotor}}$ [\si{\kg\per\m\cubed}] & 7850 \\
Young's modulus $E$ [\si{\giga\pascal}]           & 210 \\
Poisson's ratio $\nu$              & 0.29 \\
Stack mass $m_{\text{stack}}$ [\si{\kg}] & 10 \\
\hline
\end{tabular}
\end{table}

Once the finite element system is assembled, the following system of differential equations is obtained:
\begin{equation}
    \bm{M} \ddot{\bm{q}} + \bm{C} \dot{\bm{q}} + \bm{K} \bm{q} = \bm{F}_{\text{mag}}(\alpha)
\end{equation}
where $\bm{M}$, $\bm{C}$, and $\bm{K}$ are the mass, damping, and stiffness matrices, respectively, $\bm{q}$ is the vector of nodal degrees of freedom,
and $\bm{F}_{\text{mag}}$ is the electromagnetic force vector.
A modal approach is employed to reduce the dimensionality of the problem.
Resonance frequencies are reported in table \ref{table_freqTors}.
A modal damping of 4\% is introduced to attenuate the free response. The fully coupled problem is then solved using a fourth-order Runge-Kutta (RK4) time integration scheme.

\begin{table}[H]
	\caption{Natural frequencies of the rotor (in Hz)}
\begin{center}
\begin{tabular}{|*{6}{c|}}
    \hline
	\textbf{f\textsubscript{1}}  & \textbf{f\textsubscript{2}}  & \textbf{f\textsubscript{3}} & \textbf{f\textsubscript{4}} & \textbf{f\textsubscript{5}} & \textbf{f\textsubscript{6}}\\
    \hline
    0  & 1163 & 2316 & 3419 & 4199 & 4865\\
	\hline
\end{tabular}\\
\end{center}\label{table_freqTors}
\end{table}

\section{Modeling of the Electromagnetic Force}

Having established the rotor mechanical response, we now detail the modeling of the electromagnetic forces,
which are both the source of the ripple and are affected by the resulting motion.

The objective is to express the magnetic force as a function of the instantaneous rotor position,
including its oscillations. Two main electromagnetic contributions exist: magnetostriction and Maxwell's forces.
However, various studies have shown that magnetostriction forces are negligible compared to Maxwell's forces in this context \cite{Fonteyn_2010, Delaere_2002}.
Therefore, our attention is focused on the latter.

A variety of methods exist for the analysis of electric machines.
More accurate approaches include Conformal Mapping~\cite{Ragot},
models based on air gap reluctance~\cite{fakamtchakoue:tel-02061254},
the combination of permeance and magnetomotive force functions~\cite{lebesnerais:tel-00348730},
and the high-fidelity Finite Element Method (FEM)~\cite{jeannerot:hal-03549368}.
For this work, the subdomain method is employed,
as it provides an excellent trade-off between computational efficiency and accuracy in resolving both spatial and temporal harmonics.

\vspace{2em}

\subsection{Subdomain Method Formulation}

\begin{figure}[H]
 	\centering
    \includegraphics{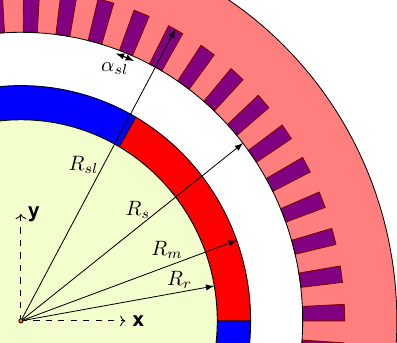}
 	\caption{Sectional view of the surface-mounted permanent-magnet machine with 54 slots and 3 pole pairs. Subdomains for the analytical model are indicated.}
    \label{f:CoupeMag}
\end{figure}

The model is based on a 2D section of a surface-mounted permanent-magnet machine, as shown in Figure \ref{f:CoupeMag}. We assume linear magnetic materials and infinitely permeable iron (saturation is neglected). The problem is solved using the magnetic vector potential $A_z$ in three subdomains: the magnets, the air gap, and the stator slots.

The governing Maxwell's equations in each subdomain are:
\begin{equation}
	\left\{
    \begin{array}{ll}
      \nabla^2 A_z = - \mu_0/r \left( M_\theta - \frac{\partial M_\theta}{\partial \theta} \right), & \text{in magnets}\\
	  \nabla^2 A_z = 0, & \text{in air gap}\\
	  \nabla^2 A_z = - \mu_0 J(t), & \text{in slots}
    \end{array}
	\right.
\end{equation}
where $J(t)$ is the current density, and $M_\theta$ is the magnetization of the magnets. The geometric and magnetic properties are listed in Table \ref{tab:em_params}. The current and rotor angle are assumed to be synchronized for maximum torque production.

\begin{table}[H]
\centering
\caption{Electromagnetic and geometric parameters of the machine.}
\label{tab:em_params}
\begin{tabular}{l r}%[table-format=2.2]
\hline
\textbf{Parameter} & \textbf{Value} \\
\hline
Rotor outer radius $R_{r}$ [\si{\mm}]       & 70    \\
Magnet inner radius $R_{m}$ [\si{\mm}]     & 61   \\
Stator inner radius $R_{s}$ [\si{\mm}]     & 71 \\
Slot outer radius $R_{sl}$ [\si{\mm}]       & 81 \\
Slot opening angle $\alpha_{sl}$ [\si{\radian}] & 0.05 \\
Number of slots                             & 54    \\
Number of poles                             & 6 \\
Current density J [\si{\ampere\per\mm\squared}] & 53.19\\
\hline
\end{tabular}
\end{table}

By applying continuity conditions at the interfaces between subdomains, we solve for $A_z$ analytically. The radial and tangential components of the magnetic flux density, $B_r$ and $B_\theta$, are then derived:
\begin{equation}
	B_r = \frac{1}{r} \frac{\partial A_z}{\partial \theta}, \quad B_{\theta} = - \frac{\partial A_z}{\partial r}
\end{equation}
The resulting flux density is expressed as a harmonic series dependent on the stator-fixed angle $\theta$ and the rotor-fixed angle $\alpha(t)$:
\begin{equation}
	\left\{
    \begin{aligned}
      & B_r(\theta, t) = \sum_{k,n} \begin{pmatrix}\sin(k\theta)&\cos(k \theta) \end{pmatrix}  \bm{B}_{r}^{(k,n)} \begin{pmatrix}\sin(n \alpha(t))\\\cos(n \alpha(t)) \end{pmatrix} \\
      & B_{\theta}(\theta, t)  = \sum_{k,n} \begin{pmatrix}\sin(k\theta)&\cos(k \theta) \end{pmatrix} \bm{B}_{\theta}^{(k,n)} \begin{pmatrix}\sin(n \alpha(t))\\\cos(n \alpha(t)) \end{pmatrix}
    \end{aligned}
	\right.
	\label{eq:ChampsB}
\end{equation}
From this, the radial and tangential Maxwell pressure $P_r = (B_r^2 - B_\theta^2)/(2 \mu_0)$ and $P_t = (B_r B_\theta)/\mu_0$ are calculated. Their general form is also a double harmonic series:
\begin{equation}
P(\theta, t) = \sum_{k,n} \begin{pmatrix}\sin(k\theta)&\cos(k \theta) \end{pmatrix}   \bm{P}^{(k,n)} \begin{pmatrix}\sin(n \alpha(t))\\\cos(n \alpha(t)) \end{pmatrix} 
\label{eq:PressionsH}
\end{equation}

\begin{figure}[H]
	\begin{subfigure}[t]{0.27\linewidth}
    \centering
    \includegraphics{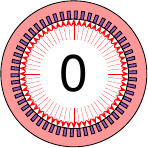}
	\end{subfigure}
	\hfill
	\begin{subfigure}[t]{0.27\linewidth}
	  \centering
    \includegraphics{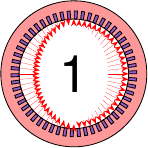}
	\end{subfigure}
	\hfill
	\begin{subfigure}[t]{0.27\linewidth}
	  \centering
    \includegraphics{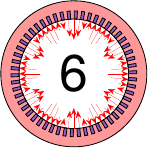}
	\end{subfigure}
	\caption{Radial pressure distribution of spatial orders 0, 1 and k=6  \label{f:SpatialOrder}}
\end{figure}

The index $k$ represents the order of the spatial distribution of the stress.
The orders 0, 1, and 6 of the radial pressure are plotted in Figure~\ref{f:SpatialOrder}.
The index $n$ represents the temporal order of the stress, i.e., how fast the preceding spatial order rotates.

\subsection{Electromagnetic Torque Expression}

\begin{figure}[H]
    \centering
    \includegraphics{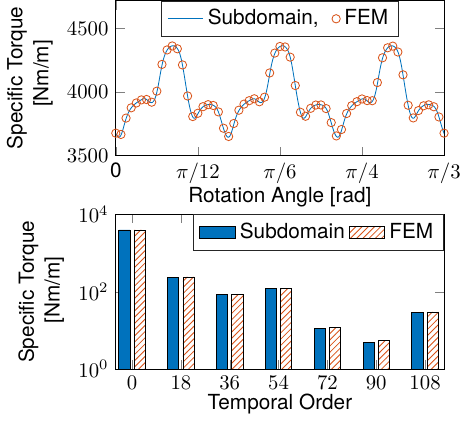}
    \caption{FEM vs. Subdomain Model Torque Comparison.}
    \label{f:Couple_FEM_SousDom}
\end{figure}

The torque is derived from Maxwell's pressures. The general expression for the torque is given by:
\begin{equation}
    \mathbf{C}_{\text{mag}}(\boldsymbol{\varepsilon},t) = \int_0^{2 \pi} R \mathbf{n} \wedge (P_r \mathbf{n} + P_\theta \mathbf{t}) \, R \, d\theta
\end{equation}
Expanding the vector product, we obtain:
\begin{equation}
    \mathbf{C}_{\text{mag}}(\boldsymbol{\varepsilon},t) = \int_0^{2 \pi} P_\theta \, R^2 \, d\theta \, \mathbf{z}
\end{equation}
The integration involves considering only $k = 0$. Thus, the expression for the magnetic torque can be written as:
\begin{equation}
    \mathbf{C}_{\text{mag}}(\boldsymbol{\varepsilon},t) = \sum_{m} 2 \pi R^2 \mathbf{P}_{k=0}^{n} \begin{pmatrix} \sin(n \alpha) \\ \cos(n \alpha) \end{pmatrix}
\end{equation}
This is compared with a FEM simulation performed using JMAG, as shown in Figure~\ref{f:Couple_FEM_SousDom}. The torque is periodic with respect to the angle, exhibiting typical multiple orders of 18, characteristic of a six-pole three-phase full-pitch winding. A very small error confirms the validity of the model.
\vspace{1cm}

\subsection{Impact of Positional Oscillations on Maxwell Pressure}

The key element of the magneto-mechanical coupling lies in how the rotor angular oscillations modulate the magnetic field. The instantaneous angle is no longer linear in time but contains oscillatory components:
\begin{equation}
	\alpha(t) = \Omega t +  \Lambda_{n_e} \cos( n_e \Omega t + \phi_{n_e})
\end{equation}
Substituting this expression into the pressure formulation (Eq. \ref{eq:PressionsH}) and applying the Jacobi-Anger expansion reveals that each original spatio-temporal harmonic generates an infinite series of sidebands. The complex amplitude $\tilde{P}_{\{k,n,m\}}$ of each new harmonic is a function of the original amplitude $P^{(0)}_{\{k,n\}}$ and a Bessel function of the first kind, $J_m$:
\begin{equation}
\tilde{P}_{\{k,n,m\}} = P^{(0)}_{\{k,n\}} \, (-i)^m \, J_m(n \Lambda_{n_e}) \, e^{i m \phi_{n_e}}
\end{equation}

In the context of electrical machines, the torque ripple and the resulting angular oscillation amplitude $\Lambda_{n_e}$ are typically small.
% This implies that the argument of the Bessel function, $z = n \Lambda_{n_e}$, is also small ($z \ll 1$), which allows for a significant simplification using the following approximations:
The Taylor development of the Bessel function leads us 
\begin{align}
    J_0 &\approx 1 - \frac{n^2 \Lambda_{n_e}^2}{4} \approx 1 \\
    J_{\pm 1} &\approx \pm \frac{n \Lambda_{n_e}}{2} \\
    J_{\pm 2} &\approx \frac{n^2 \Lambda_{n_e}^2}{8}
\end{align}
Applying these approximations to the amplitudes of the pressure harmonics yields a clear hierarchy:

\textbf{Fondamental harmonic ($m=0$):} The amplitude is almost unchanged.
    \begin{equation}
        |\tilde{P}_{\{k,n,0\}}| \approx |P^{(0)}_{\{k,n\}}|
    \end{equation}
    
\textbf{First-order sidebands ($m=\pm 1$):} These are the dominant new harmonics, with an amplitude directly proportional to the oscillation amplitude $\Lambda_{n_e}$.
    \begin{equation}
        |\tilde{P}_{\{k,n,\pm 1\}}| \approx |P^{(0)}_{\{k,n\}}| \cdot \frac{n \Lambda_{n_e}}{2}
    \end{equation}
    
\textbf{Second-order sidebands ($m=\pm 2$):} The amplitude is proportional to the \textit{square} of the oscillation amplitude, making them significantly smaller and often negligible.
    \begin{equation}
        |\tilde{P}_{\{k,n,\pm 2\}}| \approx |P^{(0)}_{\{k,n\}}| \cdot \frac{(n \Lambda_{n_e})^2}{8}
    \end{equation}

This crucial result shows that a mechanical oscillation of frequency $n_e\Omega$ primarily creates a pair of new sideband pressure waves at frequencies
$(n \pm n_e)\Omega$ around each original electrical harmonic $n\Omega$.

\section{Coupled Magneto-Mechanical Analysis}

\begin{figure}[H]
    \centering
    \includegraphics{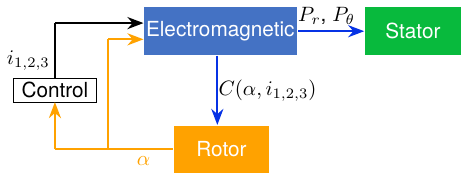}
	  \caption{Block diagram of the strongly coupled magneto-mechanical simulation.}
    \label{fig:coupling_loop}
\end{figure}

To investigate the interaction between electromagnetic forces and mechanical dynamics, the two models are integrated into a single, coupled simulation loop, as depicted in Figure~\ref{fig:coupling_loop}.

A key assumption in this model is the implementation of an ideal current controller.
This implies that the stator currents are perfectly and instantaneously synchronized with the actual rotor position,
effectively creating a direct dependency of the electromagnetic excitation on the mechanical angle.

The primary objective is to quantify the impact of torsional vibrations on the air-gap magnetic pressures. To achieve this, the coupled magneto-mechanical simulation is first executed to solve for the rotor's instantaneous angle. This dynamic angle is then fed back into the electromagnetic model to calculate the resulting Maxwell stress tensor under realistic operating conditions.

%\vspace{2em}

\subsection{Rotor Response in Coupled Simulation}
$\,$
\vspace{-2.5em}
\subsubsection{Run-up Simulation}

\begin{figure}[H]
    \centering
    \includegraphics{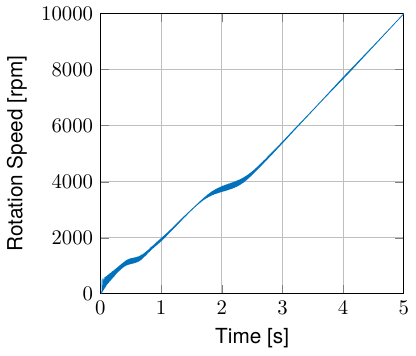}
    \caption{Simulated rotor speed during run-up, showing resonance plateaus.}
    \label{f:TorsionRunUp}
\end{figure}

Figure \ref{f:TorsionRunUp} illustrates the simulated run-up of the rotor.
Plateaus in the rotation speed curve are visible when the excitation frequency of a major torque harmonic (or one of its multiples) coincide a torsional natural frequency of the rotor.
This corresponds to a resonant amplification of the torsional vibrations.
As predicted by Equation \eqref{eq:alpha_taylor_split}, this increased oscillation leads to a slight decrease in the mean electromagnetic torque, thus momentarily halting acceleration.

\begin{figure}[H]
	\centering
    \includegraphics{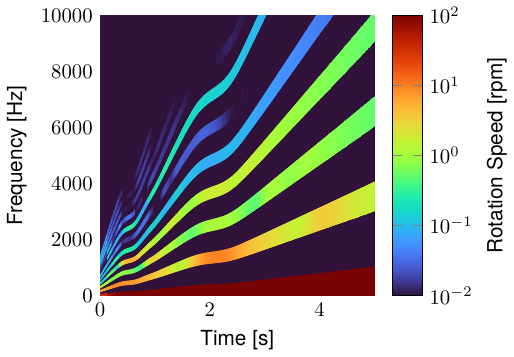}
	\caption{Spectrogram of the rotor angular acceleration during run-up.}
    \label{f:Spectro} 
\end{figure}

The accompanying spectrogram (Figure \ref{f:Spectro}) confirms that the excitation spectrum is rich, containing not only the fundamental electrical harmonics but also their multiples, which are responsible for exciting the different torsional modes.

\subsubsection{Steady-State Regime}
\begin{figure}[H]
	\centering
    \includegraphics{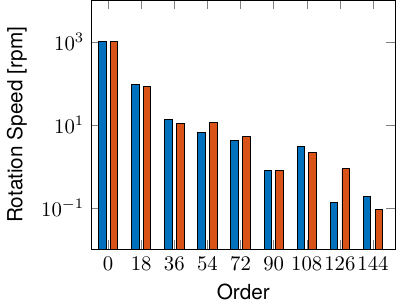}
	\caption{FFT of velocity ripple, showing amplitude differences between coupled (orange) and uncoupled model (blue).} 
	\label{f:1000_b}
\end{figure}

% well below the first torsional resonance,
At a steady-state speed of \SI{1000}{\rpm}, the system's behavior is governed primarily by the rigid body mode.
We compare the velocity ripple from the fully coupled model with that
from a simplified approach where the torque ripple is treated as a purely time-dependent excitation (i.e., ignoring the $\cos(p\alpha(t))$ dependency).
\begin{figure}[H]
    \centering
    \includegraphics{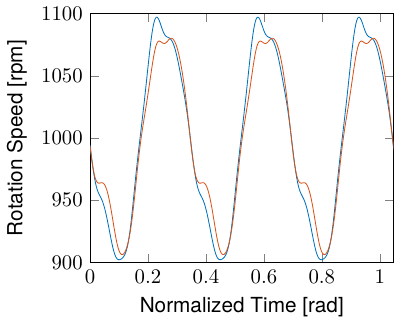}
    \caption{Comparison of velocity ripple at \SI{1000}{\rpm}: Coupled (orange) vs. Uncoupled model (blue).}
    \label{f:1000_a}
\end{figure}
While the time-domain velocity ripple appears similar (Figure \ref{f:1000_a}), the frequency domain analysis (Figure \ref{f:1000_b}) reveals
notable differences in the amplitude of certain harmonics, highlighting the inaccuracy of the uncoupled model even in non-resonant conditions.
However, the overall impact on the rotor kinematic performance and efficiency is minor.

\subsection{Impact on Maxwell Pressures}

\begin{figure}[H]
    \centering
    \includegraphics{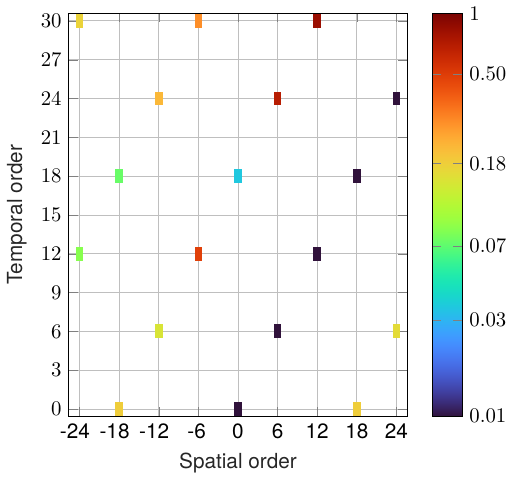}
	\caption{Relative difference in radial pressure harmonics between weak and strong coupling models at \SI{1000}{\rpm}.}
    \label{f:PressionsFortTors}
\end{figure}

The most significant consequence of this coupling is observed in the Maxwell pressure distribution on the stator.
Figure \ref{f:PressionsFortTors} shows the relative difference in the spatio-temporal spectrum of the radial pressure between the uncoupled (weak coupling)
and the fully coupled (strong coupling) simulations at \SI{1000}{\rpm}.

The plot clearly shows that, in the coupled simulation, the amplitudes of existing harmonics are modified by a non-negligible modulation.
These modifications to the pressure field are critical as they represent the excitation source for stator vibrations and, consequently, acoustic noise. An uncoupled model would fail to predict these pressure components, potentially leading to an inaccurate NVH assessment of the machine.

\section{Conclusion}

This paper investigated the closed-loop interaction between electromagnetic torque ripple and the mechanical dynamics of a rotor. We have demonstrated, both analytically and through coupled simulation, that the common assumption of a constant rotational speed is an oversimplification that neglects important physical phenomena.

The key findings are as follows:
\begin{enumerate}
    \item The rotor response to torque ripple is inherently non-linear, leading to the generation of harmonics in its angular motion and a slight decrease in its mean speed.
    \item These mechanical oscillations modulate the air-gap magnetic field.
    Through a mechanism described by Bessel functions, this modulation creates new spatio-temporal harmonics in the Maxwell pressure distribution.
    \item A strongly coupled magneto-mechanical model reveals that these new modulated pressure harmonics, not predicted by conventional models,
    can reach significant amplitudes, especially when a torsional mode of the rotor is excited.
\end{enumerate}

This finding is crucial for high-fidelity Noise, Vibration, and Harshness (NVH) analysis of electrical machines.
The previously unmodeled pressure harmonics can act as new excitation sources that may coincide with structural modes of the stator,
contributing to the machine's overall acoustic signature.

\printbibliography

\end{multicols}
\end{document}